# Superconductivity and magnetism in compressed actinium-beryllium-hydrogen alloys

Zhongyu Wan[1], Tianyi Yang[1], Wenjun Xu[1], Ruiqin Zhang[1][2]*

[1] Department of Physics, City University of Hong Kong, Hong Kong SAR 999077, People's Republic of China

[2] Beijing Computational Science Research Center, Beijing, 100193, People's Republic of China

*E-mail: aprqz@cityu.edu.hk

**Abstract**

The discovery of new high-temperature superconductors is one of the most critical problems in materials, chemistry, and physics. This work systematically investigates fluorite-like structures of $XBeH_8$ (X=Ac,Th,Pa,U, and Np) to gain physical insight into the pressure effects on the properties. Our results reveal that $AcBeH_8$ and $ThBeH_8$ are two potential high-temperature superconductors, where $AcBeH_8$ reaching a superconducting transition temperature ($T_c$) of 284.11 K at 150 GPa and even a value of 203.29 K at 50 GPa, $ThBeH_8$ has $T_c$=217.65 K at 200 GPa, and the predicted results provide strong candidates for achieving high-temperature or even room-temperature superconductivity. The analysis of pressure effects demonstrates unusual charge transfer and atomic structure in the system, with $s \rightarrow f$ migration in Ac/Th atoms and $s \rightarrow p$ migration in Be atoms, similar to the high-pressure behavior of alkaline earth metals. In the past long time, superconductivity and magnetism were considered unable to coexist. However, $UBeH_8$ and $NpBeH_8$ are two potential magnetic superconductors with maximum $T_c$ of 60.70 and 107.18 K at 100 and 350 GPa, respectively, and Hubbard U calculations predict their magnetic moments with 1.52-1.90 and 2.49-2.92 $\mu_B$ at 0 K. The coexistence of magnetic and superconducting states may be attributed to the different sublattice actions.

# 1. Introduction

Since superconductors exhibit zero resistance in the superconducting state,[1] they can potentially be used in high-voltage transmission lines.[2] However, it will transform into a superconductor only when it is below the superconducting transition temperature ($T_c$),[3] so $T_c$ is one of the most critical properties determining superconductors' application. For most superconductors, such as copper-based,[4] iron-based,[5] and $MgB_2$,[6] their $T_c$ can reach 133, 56, and 39 K. The relatively low $T_c$ makes their application in practice limited. Therefore, the search for superconductors with high $T_c$ is currently one of the most critical projects in physics, materials, and chemistry.

The Bardeen-Cooper-Schrieffer (BCS) theory[7] claims that the $T_c$ of conventional superconductors is positively correlated with the Debye temperature ($T_D$), and $T_D$ is inversely proportional to the atomic mass. So metallic hydrogen at high pressure is predicted to be an ideal room temperature superconductor,[8] and it is experimentally observed at 495 GPa.[9] Despite having satisfactory $T_c$, however, the extremely high pressure makes it difficult to synthesize and preserve. Ashcroft et al.[10] proposed a "chemical precompression" strategy to achieve metallization at lower pressures by secondary compression of the hydrogen sublattice with other elements. Compared to metallic hydrogen, these hydrogen-rich compounds maintain a relatively high $T_c$ and lower synthetic pressure,[11] regarded as potential candidates for achieving high-temperature or even room-temperature superconductivity.[12] There are two classical types of hydrogen-rich compounds, one is the covalent hydrides $H_3S$ (155GPa, $T_c$=203K),[13] $SiH_4$ (202GPa, $T_c$=166K),[14] $PH_2$ (220GPa, $T_c$=78K),[15] and the other is the hydrogen cage compounds $LaH_{10}$ (170GPa, $T_c$=250K),[16] $CaH_6$ (172GPa, $T_c$=215K),[17] $ThH_{10}$ (100GPa, $T_c$=241K),[18] $AcH_{10}$ (200GPa, $T_c$=251K),[19] and $YH_6$ (183GPa, $T_c$=220K)[20] with metal elements. In contrast, the $T_c$ of the latter one is much closer to room temperature, especially for the lanthanide and actinide hydrides.

It is worth noting that adding another element to the binary hydride may enhance the "chemical precompression" effect. Recently, more and more ternary hydrogen-rich compounds are predicted to be capable of high-temperature superconductivity at relatively low pressures, such as $KB_2H_8$ (12GPa, $T_c$=146K),[21] $LaBeH_8$ (20GPa, $T_c$=185K),[22] and $LaBH_8$ (50GPa, $T_c$=126K).[23] Especially in the fluorite-like structure of the $LaBeH_8$ and $LaBH_8$ compounds, there is a sublattice of hydrogen cages, which may be the reason for its high $T_c$ at low pressure. Therefore, it is reasonable to believe that actinide ternary hydrides with the same structure are also potential high-temperature superconductors.

Based on the above considerations, the fluorite-like structure of XBeH$_8$ compounds (X = Ac, Th, Pa, U, and Np) are chosen to do a systematic study to discover potential high-$T_c$ superconductors, as well as obtaining physical insight into the superconductivity depended on different pressures and elements. This work provides theoretical guidance for the discovery and pressure modulation of high-temperature superconductors.

## 2. Computational Details

The CALYPSO software[24] is used to perform a search for possible XBeH$_8$ (X = Ac, Th, Pa, U, and Np) crystal structures from 0 to 500 GPa. The Vienna Ab-initio Simulation Package (VASP)[25] is used to perform density functional calculations, and the projected-augmented wave pseudopotentials method[26] is employed to describe the electron-ion interaction. Generalized gradient approximation of Perdew-Burke-Ernzerhof theory as exchange-correlated functional[27] are employed with a cutoff energy of 900 eV and k-point grid density of 0.03 × 2π Å$^{-1}$ based on the Monkhorst-Pack method[28]. $6s^26p^66d^17s^2$, $6s^26p^66d^27s^2$, $6s^26p^65f^26d^17s^2$, $6s^26p^65f^36d^17s^2$, $6s^26p^65f^46d^17s^2$, $2s^2$, and $1s^1$ are treated as the valence electron structures of Ac, Th, Pa, U, Np, Be, and H respectively. The relaxed ionic positions and cell parameters provided energy and forces smaller than 10$^{-5}$ eV and 0.003 eV/Å, respectively. The QUANTUM ESPRESSO package (QE)[29] is used to calculate the electron-phonon coupling (EPC) properties in linear response theory, the ultra-soft pseudopotential with a kinetic cutoff energy of 150 Ry and a charge density cutoff energy of 1500 Ry, respectively. Self-consistent field (scf) calculations are performed on a k-point grid with a density of 16 × 16 × 16, the Methfessel-Paxton first-order spread is set to 0.02 Ry, and an irreducible q-point grid with a phonon density of 4 × 4 × 4 is used to further calculate the phonon property. The double $\delta$ function for phonons are smeared out with widths of 0.5 meV, the EPC coefficient ($\lambda$) and $T_c$ are obtained by solving anisotropic Eliashberg equations (Coulomb pseudopotential $\mu^*$=0.1)[30] with the Electron-Phonon coupling using Wannier functions (EPW) package[31, 32].

## 3. Results and Discussion

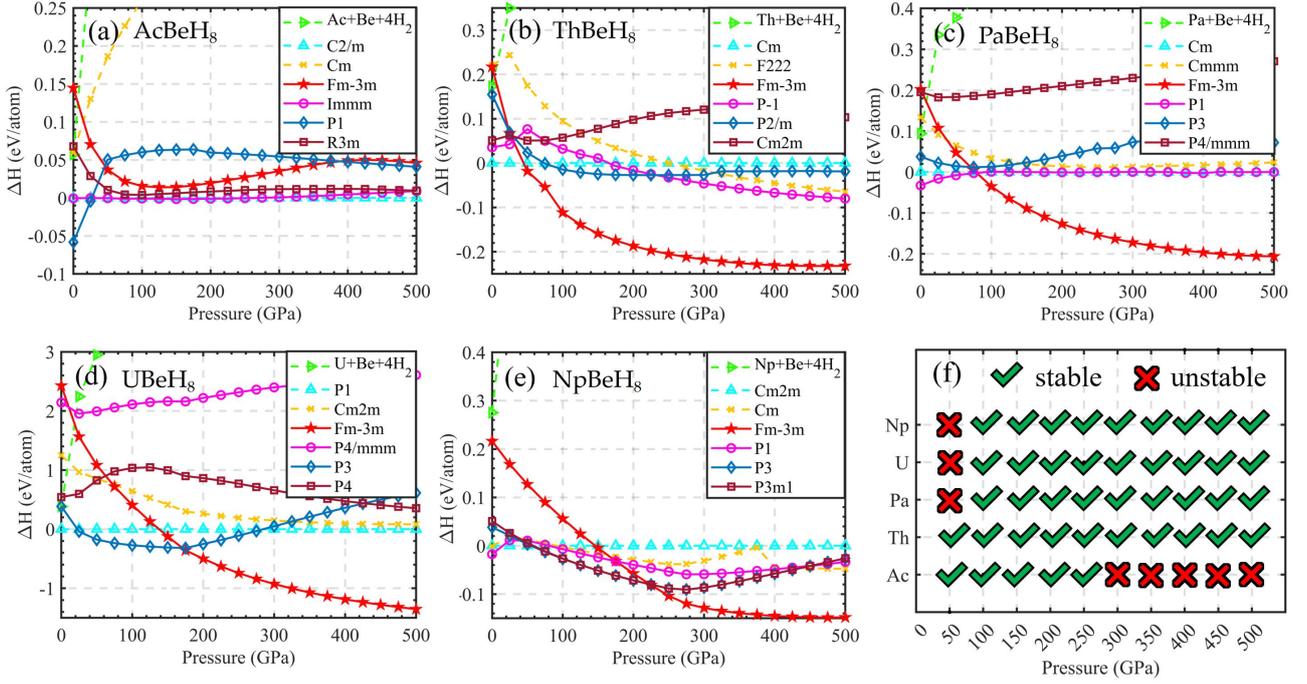

**Figure 1. Thermodynamic and kinetic stability of compounds. (a)-(e)** The enthalpies of XBeH$_8$ (X = Ac, Th, Pa, U, and Np) at pressures from 0 to 500 GPa and considering their decomposition into the monomers Ac,[33] Th,[34] Pa,[35] U,[36] Np,[37] Be,[38] and H.[39] **(f)** Kinetic stability of XBeH$_8$ per 50 GPa.

The stability of a compound is crucial in determining whether it can be synthesized experimentally. Hence, the first step in our work is to determine these compounds' thermodynamic and kinetic stability. From Figures. 1(a)-(e), it can be seen that ThBeH$_8$, PaBeH$_8$, UBeH$_8$, and NpBeH$_8$ with fluorite-like structures (space group: $Fm\bar{3}m$) are thermodynamically most stable at pressures above 50, 100, 175, and 225 GPa, respectively. Moreover, Th and Pa, U, Np can also be kinetically stable in the pressure range of 50-500 GPa and 100-500 GPa. Unfortunately, AcBeH$_8$ is thermodynamically metastable and can remain kinetically stable at 50-250 GPa. In fact, conditionally stable phases (thermodynamically metastable, kinetically stable) can also be found experimentally and even dominate over thermodynamically stable phases.[40] Therefore, we cannot deny that it can be synthesized from experiments.

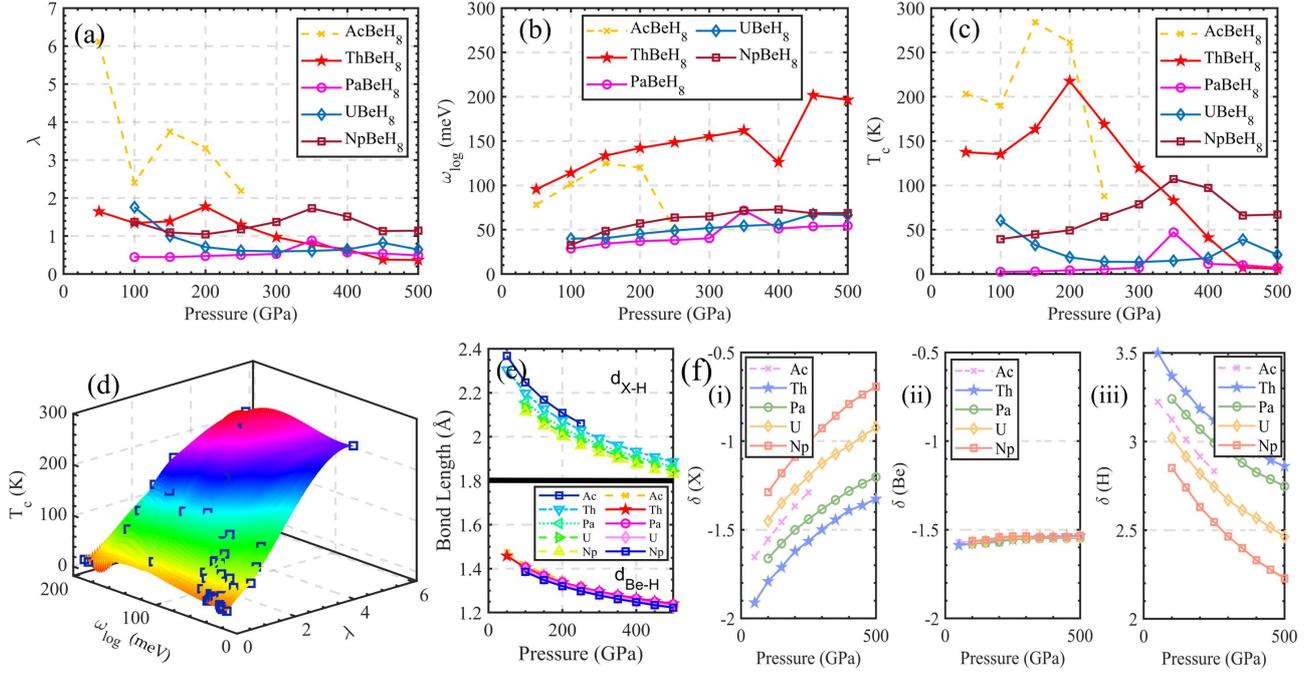

**Figure 2.** (a)-(c) $\lambda$, $\omega_{\log}$, and $T_c$ under pressure induction. (d) Three-dimensional surfaces of $T_c$ versus $\lambda$ and $\omega_{\log}$. (e) Bond lengths of X-H and Be-H under pressure induction. (f) Pressure-induced Bader charge of elements per unit cell in different system. (i) X atom. (ii) Be atom. (iii) H atoms.

Kinetic stability is a prerequisite for $T_c$ prediction. Based on the results in Figure 1(f), $\lambda$, log-averaged phonon frequency ($\omega_{\log}$), and $T_c$ are calculated and presented in Figure 2. It reveals that $AcBeH_8$ has stronger electron-phonon interactions than the other four systems, even with $\lambda$= 6.12 at 50 GPa, which is much higher than that of $LaBeH_8$[22] ($\lambda$= 2.19) and $LaBH_8$[23] ($\lambda$= 2.29), implying that $AcBeH_8$ at 50 GPa could be potentially an excellent superconductor. Compared to the elements Th, Pa, U, and Np. Ac has a larger atomic radius,[41] which means that the pressure will subject Ac to greater Coulomb repulsion, leading to metastability from the Jahn-Teller (JT) effect,[42] and thus, this may account for the metastability of $AcBeH_8$. Furthermore, the JT effect can enhance the electron-phonon interaction,[43] it explains the high $\lambda$ of $AcBeH_8$. Surprisingly, $AcBeH_8$ at 50 GPa has $T_c$=203.29 K and $T_c$=284.11 K at 150 GPa, which is very close to room temperature. Its strong EPC may cause it. While $ThBeH_8$ can reach a $T_c$ of 217.65K at 200 GPa, which may be attributed to its high $\omega_{\log}$. Figure 2(d) depicts the relationship between $T_c$, $\lambda$, and $\omega_{\log}$. Having both high $\lambda$ and $\omega_{\log}$ may be the possible reason for an excellent $T_c$.

Pressure affects the crystal parameters, and Figure 2(e) shows the variation of X-H bond length ($d_{X-H}$) and Be-H bond length ($d_{Be-H}$) with pressure. Regardless of the $XBeH_8$ system, increasing pressure causes continuous shortening of $d_{X-H}$ and $d_{Be-H}$. The bond length affects the charge transfer between atoms and electronic distribution, so Bader charge analysis[44] is used to describe this physical process quantitatively. It can be seen from Figure 2(f) that regardless of the elements of X, the charge of Be atoms is minimally perturbed by the pressure. In contrast, X and H atoms have significant charge changes, which may be since Be atoms have higher ionization energy (899.5 kJ/mol). At the same time, Ac-Np's is only 499-604.5 kJ/mol,[45] and thus Be atoms are more strongly bound to valence electrons and uneasy to be affected by H atoms. This figure also shows that the increased pressure makes the X atom provide fewer electrons to H atoms.

In contrast to the usual reduction in atomic spacing, which enhances charge transfer, this anomalous result shows that this system may have a unique electron distribution. Considering that the calculated radii of Ac, Th, Pa, U, Np, and H are 5.30, 5.06, 3.70, 3.22, 2.84, and 0.53 Å.[41] But $d_{\text{X-H}}$ is much smaller than the sum of two atomic radii. It is the great Coulomb repulsion that prevents the charge transfer.

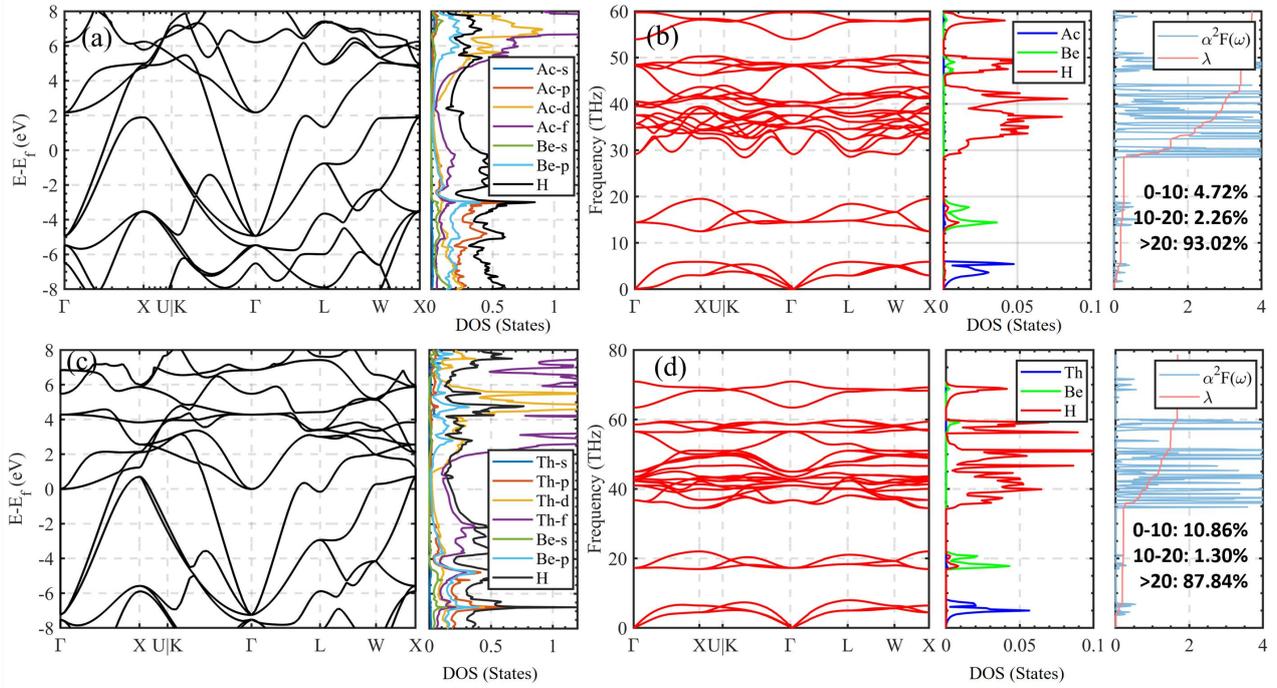

**Figure 3.** (a) Band structure and electronic density of states (DOS) of AcBeH$_8$. (b) Phonon dispersion curves, Phonon DOS and Eliashberg spectral functions of AcBeH$_8$. (c) Band structure and electronic DOS of ThBeH$_8$. (d) Phonon dispersion curves, PhDOS and Eliashberg spectral functions of ThBeH$_8$.

Both AcBeH$_8$ (150 GPa) and ThBeH$_8$ (200 GPa) have outstanding $T_c$, so it is necessary to investigate its mechanistic origin. Properties of the solid depend on its electronic structure,[46] as shown in Figure 3. Strikingly, the DOS of Ac and Th have certain $f$ electrons below the Fermi level and $p$ electrons present for Be atoms, and the $s$-electrons of Ac and Th are not observed in DOS. However, $f$ and $p$ electrons don't exist in their ground state atomic structures, respectively. The emergence of novel electronic states is most likely pressure-induced due to the dense overlap of electron clouds making the energy of atomic electrons increase. The energy will determine the orbital occupation of electrons in the atomic structure. Thus, the large interatomic repulsive forces may change the occupied electronic states, resulting in a redistribution of electrons. This is similar to the behavior of alkaline earth metals Na, K, Rb, and Cs at high pressure.[47, 48] Our results reveal a possible intra-atomic $s \rightarrow f$ migration for Ac and Th, and a $s \rightarrow p$ migration for Be. Besides, Ac, Th have the same trend as Be, H on DOS, respectively, which implies a strong coupling between these atoms, which are consistent with the characteristics of high-temperature superconductors.[12] Figures 3 (b),(d) reveal the phonon vibrations in the two superconductors, with a clear regionalization of the phonon density of states for different elements, where the low-frequency phonons at 0-10 THz are mainly occupied by Ac/Th, while the medium-frequency phonons at 10-20 THz are mainly contributed by Be, while the high-frequency phonons above 20 THz are provided by H atoms, due to the fact that atoms with smaller masses are subject to larger force constants and have stronger vibrational frequencies. The Eliashberg spectral function reveals that the contribution of high-frequency phonons to $\lambda$ is 93.02% and 87.84%, respectively, which far exceeds that of the medium and low frequencies, while the high-frequency phonons are mainly from H atoms, which dominate EPC interactions.

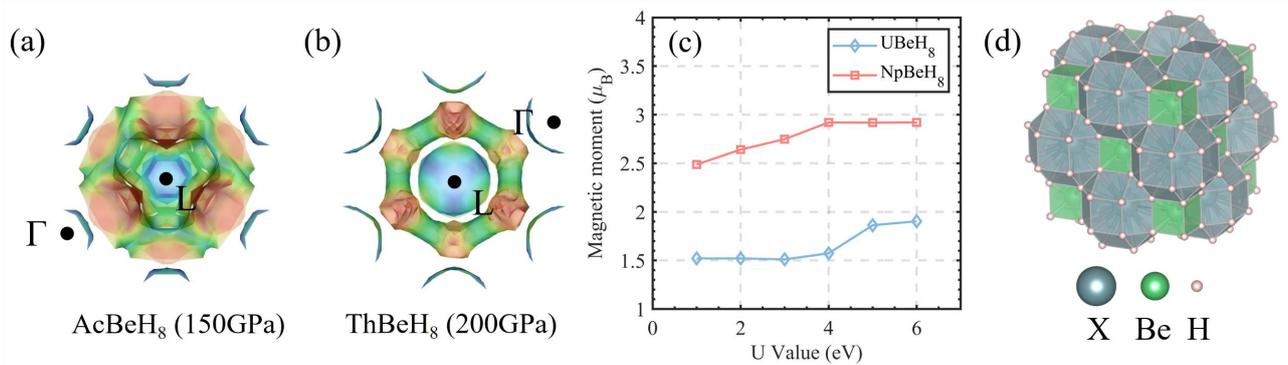

**Figure 4. (a) Fermi surface of AcBeH$_8$ at 150 GPa. (b) Fermi surface of ThBeH$_8$ at 200 GPa. (c) Magnetic moments of UBeH$_8$ and NpBeH$_8$ with different U values at 100 and 350 GPa, respectively. (d) The fluorite-like crystal structure of XBeH$_8$.**

The high-temperature superconductivity may also be related to the behavior of Fermi surfaces.[49] Figure 4(a), (b) shows the Fermi surfaces of AcBeH$_8$ and ThBeH$_8$ at the pressures corresponding to the highest $T_c$. It can be clearly seen that there is Fermi surface nesting from the $\Gamma \rightarrow L$ direction, which can enhance electron pairing to form Cooper pairs.[50] Their superconducting gaps are 43.09 and 33.01 meV, respectively. The high superconducting gaps mean that their Cooper pairs are difficult to be disassembled. The maximum $T_c$ of PaBeH$_8$, UBeH$_8$, and NpBeH$_8$ is 47.05, 60.70, and 107.18 K, which indicates that they also have certain superconductivity at high pressure. Notably, the hydrides of U and Np are magnetic,[51, 52] which implies that UBeH$_8$ and NpBeH$_8$ may be potential magnetic compounds. In order to accurately describe the nature of magnetism, the 5$f$ electrons of U and Np are treated by the Hubbard U model.[53] In this approach, U is an empirical parameter, and different values of U affect the magnetic results,[54] however the systems concerning U-Be-H and Np-Be-H remain experimentally unreported, so U = 1, 2, 3, 4, 5, 6 eV is employed to cover a wide range of possibilities. At the pressure corresponding to the maximum $T_c$, our DFT+U calculations explain that UBeH$_8$ and NpBeH$_8$ are indeed magnetic compounds. Their magnetic moments at 100 and 350 GPa are 1.52-1.90 and 2.49-2.92 $\mu_B$, respectively, with magnetic moments contributed all by U/Np. However, superconductors were considered completely antimagnetic (Meissner effect) for a long time,[55] but our results demonstrate novel states of superconductivity and magnetic coexistence in UBeH$_8$ and NpBeH$_8$. Based on the above results, the source of magnetism in the system is U/Np, however, the main contributor to the EPC interaction is the cage sublattice of hydrogen, where magnetism and superconductivity operate in their respective sublattices and are weakly associated with each other (Figure 4(d)), thus suppressing the mutual repulsion between superconductivity and magnetism. However, their superconducting gaps are 9.21 (UBeH$_8$) and 16.26 meV (NpBeH$_8$), respectively, and their relatively low gaps mean that Cooper pairs that tend to form superconducting states are easily destroyed by magnetic particles in the system, which provides an explanation for their having $T_c$ but not high.

## 4. Conclusion

This work systematically investigates the fluorite-like structure of XBeH$_8$ (X = Ac, Th, Pa, U, and Np). EPC calculations reveal several potential high-temperature superconductors, the conditionally stable AcBeH$_8$ has a $T_c$ of 284.11 K at 150 GPa. Even it can reach a $T_c$ of 203.29 K at 50 GPa, the JT effect is a possible reason for the system's high $\lambda$. The $T_c$ of ThBeH$_8$ at 200 GPa is 217.65 K. This provides a strong candidate for achieving high-temperature or even room-temperature superconductivity. Our results also reveal an unusual charge transfer at high pressures for XBeH$_8$, where the extremely short interatomic distances elevate their electron energies, leading to a redistribution of electrons, with $s \rightarrow f$ migration for Ac and Th, and the presence of $s \rightarrow p$ migration for Be. Further DFT+U calculations reveal that UBeH$_8$ and NpBeH$_8$ are magnetic superconductors with $T_c$ of 60.70, 107.18 K and magnetic moments of 1.52-1.90, 2.49-2.92 $\mu_B$ at 100 and 350 GPa, respectively, and the coexistence of magnetic and superconducting states may be attributed to the separate effects of different sublattices. This work provides theoretical guidance for the design of new high-temperature superconductors and novel states of matter.

**Declaration of Competing Interest**

The authors declare that they have no known competing financial interests or personal relationships that could have appeared to influence the work reported in this paper.

**Acknowledgements**

The work described in this paper was supported by grants from the Research Grants Council of the Hong Kong SAR (CityU 11305618 and 11306219) and the National Natural Science Foundation of China (11874081).

**Supplementary materials**

Supplementary materials to this article can be found online.